\documentclass{ws-ijmpa}

\begin{document}

\markboth{F. Cianfrani, G. Montani}
{The role of the time gauge in the 2nd order formalism}

%
\catchline{}{}{}{}{}
%

\title{The role of the time gauge in the 2nd order formalism
}

\author{FRANCESCO CIANFRANI}

\address{ICRA---International Center for Relativistic Astrophysics\\ 
Dipartimento di Fisica (G9),
Universit\`a  di Roma, ``Sapienza'',\\
Piazzale Aldo Moro 5, 00185 Rome, Italy.\\ 
francesco.cianfrani@icra.it}

\author{GIOVANNI MONTANI}

\address{ICRA---International Center for Relativistic Astrophysics\\ 
Dipartimento di Fisica (G9),
Universit\`a  di Roma, ``Sapienza'',\\ 
Piazzale Aldo Moro 5, 00185 Rome, Italy.\\ 
ENEA C.R. Frascati (Dipartimento F.P.N.),
Via Enrico Fermi 45, 00044 Frascati, Rome, Italy.\\
ICRANet C. C. Pescara, Piazzale della Repubblica, 10, 65100 Pescara, Italy.\\
montani@icra.it}

\maketitle

\begin{history}
\received{Day Month Year}
\revised{Day Month Year}
\end{history}

\begin{abstract}

We perform a canonical quantization of gravity in a second-order formulation, taking as configuration variables those describing a 4-bein, not adapted to the space-time splitting. We outline how, neither if we fix the Lorentz frame before quantizing, nor if we perform no gauge fixing at all, is invariance under boost transformations affected by the quantization.


\end{abstract}

\ccode{PACS number: 04.60.-m}

\section{Introduction}
The Loop Quantum Gravity formulation (for a review see \cite{th}) is based on fixing, before quantizing, the so-called time-gauge condition, which corresponds to set the 4-bein vectors such that the time-like one $e_0$ is normal to spatial hypersurfaces. 
If this hypothesis is neglected, a deep complication occurs in a first order formulation, since second-class constraints arise. 
Barros and Sa \cite{BS} demonstrated that these second-class constraints can be solved, such that only first class ones remain. He concluded that this feature demonstrate that boosts can be safely fixed before the quantization 

However, Alexandrov \cite{Al,AL} provided a covariant formulation in which second-class constraints are solved by replacing Poisson brackets with Dirac ones. This way he was able to recognize a Gauss constraint for the Lorentz group, but the discrete structure is lost. 

Therefore, the development of a formulation, in which the Lorentz frame is not fixed, can provide a deep insight towards the understanding of gravitational quantum features.

We are going to provide the proof that in a second order formulation of General Relativity, the time-gauge condition does not provide any boost symmetry violation on a quantum level \cite{CM07}. This result has been obtained by fixing the Lorentz frame before the quantization and recognizing a unitary operator connecting the quantum description in different frames.

\section{Hamiltonian formulation without the time gauge}

Let us consider an hyperbolic space-time manifold $V$ endowed with a metric $g_{\mu\nu}$ and a $3+1$ representation 

\begin{equation} 
V\rightarrow\Sigma\otimes \mathcal{R}, 
\end{equation}

$\Sigma$ being spatial 3-hypersurfaces with coordinates $x^i\hspace{0.2cm}(i=1,2,3)$, while 
$t$ denotes the real (time-like) coordinate. The vector field $n$ normal to $\Sigma$ can be written in terms of the lapse function $\tilde{N}$ and the shift vector $\tilde{N}^i$ as follows

\begin{equation}
n=\tilde{N}dt+\tilde{N}_idx^i.
\end{equation}

If 4-bein vectors $e^A$ are introduced, they can be expressed as

\begin{equation}
  e^0=Ndt+\chi_a E^a_idx^i,\qquad e^a=E^a_iN^idt+E^a_idx^i,
\end{equation} 

where the links between their components and the lapse function, the shift vector and the 3-metric $h_{ij}$ are given by the following relations
 
\begin{eqnarray} 
\tilde{N}=\frac{1}{\sqrt{1-\chi^2}}(N-N^iE_i^a\chi_a)\qquad
\tilde{N}^i=N^i+\frac{E^c_l\chi_cN^l-N}{1-\chi^2}E^i_a\chi^a\\
h_{ij}=E^a_iE^b_j(\delta_{ab}-\chi_a\chi_b).
\end{eqnarray}

3-bein vectors corresponding to $h_{ij}$ read
$E'^a_i=E^b_i(\delta^a_b-\alpha\chi^a\chi_b)$ with $\alpha=(1-\sqrt{1-\chi^2})/\chi^2$.

The time-gauge condition means setting $e_a$ on $\Sigma$ and it can be obtained by a boost $e'^B=\Lambda^B_{\phantom1A}e^A$ with $\chi_a=-\Lambda^0_{\phantom1a}/\Lambda^0_{\phantom10}$.

Therefore if $\chi_a\neq 0$ the $e^A$ frame is not at rest with respect to $\Sigma$, but moves with velocity components $\chi^a$. We expect that some device (classical matter field) allows to give a gauge invariant character to the statement ``at rest with respect to  $\Sigma$''.

Let us now consider a Lagrangian formulation taking as configuration variables $\tilde{N}, \tilde{N}^i, E^a_i, \chi_a$; by evaluating conjugate momenta ($\pi_{\tilde{N}}$, $\pi_i$, $\pi^i_a$ and $\pi^a$), one recognizes that there are some constraints \cite{CM07}. Hence, once Lagrangian multipliers $ \lambda^{\tilde{N}},\lambda^i,\lambda_a,\eta_a$ are considered, together with $ \tilde{N}'=\sqrt{h}\tilde{N}$, the Hamiltonian density can be written as

\begin{equation}
\mathcal{H}=\tilde{N}'H+\tilde{N}^iH_i+\lambda^{\tilde{N}}\pi_{\tilde{N}}+\lambda^i\pi_i+\eta_a\varphi^a+\lambda_a\Phi^a,
\end{equation}

where $H$ and $H_i$ are the super-Hamiltonian and the super-momentum, respectively, whose expressions are given by

\begin{equation}
H=\pi^i_a\pi^j_b\bigg(\frac{1}{2}E^a_iE^b_j-E^b_iE^a_j\bigg)+h{}^{3}\!R,\qquad H_i=D_j(\pi^j_aE^a_i).
\end{equation}
It is well-known that conditions $H=H_i=0$ arise as secondary constraints from primary ones $\pi_{\tilde{N}}=\pi_i=0$.

As far as $\Phi^a$ and $\varphi^a$ are concerned, their expressions are as follows 
\begin{eqnarray}
\Phi^a=\pi^a-\pi^b\chi_b\chi^a+\delta^{ab}\pi^i_b\chi_cE^c_i\\
\varphi^a=\epsilon_{\phantom1b}^{a\phantom1c}(\pi^b\chi_{c}-\pi^i_{c}E^b_i)
\end{eqnarray}
and, from their action on the phase space, $\Phi^a=0$ and $\varphi^a=0$ ensure the invariance under boosts and rotations of the 4-bein, respectively.

\section{Quantization after gauge fixing}

A Lorentz frame can be chosen by fixing $\chi_a$ equal to some space-time functions $\bar{\chi}_a(t,x^i)$ and solving boost, so finding
\begin{equation}
\pi^a=-\bigg(\delta^{ab}+\frac{\bar{\chi}^a\bar{\chi}^b}{1-\bar{\chi}^2}\bigg)\pi^i_b\bar{\chi}_cE^c_i.
\end{equation}

Hence, in a boost-fixed reference the action can be written as
\begin{eqnarray*}
  S=-\frac{1}{16\pi G}\int[\pi^i_a\partial_tE^a_i+\pi_{\tilde{N}'}\partial_t\tilde{N}'+\pi_i\partial_t\tilde{N}^i-\tilde{N}'H^{\bar\chi}-\tilde{N}^iH^{\bar\chi}_i-\lambda_a\varphi_{\bar{\chi}}^a-\lambda^{\tilde{N}}\pi_{\tilde{N}}-\lambda^i\pi_i]dtd^3x,
\end{eqnarray*}

where

\begin{eqnarray}
 \varphi_{\bar{\chi}}^a=\epsilon^{abc}(\bar{\chi}_b\pi^i_dE^d_i\bar{\chi}_d-\delta_{db}\pi^i_{c}E^d_i)
\end{eqnarray}
is the new rotation constraint.

The quantization can now be performed canonically, by promoting $E^a_i$, $\tilde{N}$ and $\tilde{N}^i$ to multiplicative operators and replacing Poisson brackets with commutators. 
 
Let us consider the sector $\bar{\chi}_a=0$, where the full set of constraints reads as
\begin{equation*}
  H^{0}\psi_0=0,\quad H^{0}_i\psi_0=0,\quad\epsilon^{ab}_{\phantom1\phantom2c}\pi^i_bE^c_i\psi_0=0.
\end{equation*}

The action of boost constraints in the $(E^a_i,\pi^j_b)$ phase space is reproduced by the unitary operator
\begin{equation*}
  U_\epsilon=I-\frac{i}{4}\int\epsilon^a\epsilon_b(E^b_i\pi^i_a+\pi^i_aE^b_i)d^3x+O(\epsilon^4),
\end{equation*}

which sends 3-bein vectors for $\bar{\chi}_a=0$ to 3-bein vectors for $\bar{\chi}_a=\epsilon_a$, {\it i.e.}

\begin{equation}
  U^\dag_\epsilon E^a_iU_\epsilon^{-1}=E^b_i(\delta^a_b-\frac{1}{2}\epsilon^a\epsilon_b)+O(\epsilon^4)={E'}^a_i+O(\epsilon^4).
\end{equation}

The transformation between $\chi$-sectors can be realized as a quantum symmetry if the unitary operator $U$ maps physical states to physical states.

The new state $\psi'=U_\epsilon\psi_0$ satisfies
\begin{equation*}
  U_\epsilon H^0U_\epsilon^{-1}\psi'=0\quad U_\epsilon H^0_iU_\epsilon^{-1}\psi'=0\quad U_\epsilon\epsilon^{ab}_{\phantom1\phantom2c}\pi^i_bE^c_iU^{-1}_\epsilon\psi'=0.
\end{equation*}
 
One finds at the $\epsilon^2$ order that

\begin{equation}
U_\epsilon H^0U_\epsilon^{-1}=H^{\epsilon}\hspace{0.4cm}U_\epsilon H^0_iU_\epsilon^{-1}=H^{\epsilon}_i,
\end{equation}

hence solutions of the super-momentum and of the super-Hamiltonian constraints for $\chi_a=0$ are mapped to solutions of the super-momentum and of the super-Hamiltonian constraints for $\chi_a=\epsilon_a$.

As far the rotation constraint is concerned, one gets
\begin{equation}
U_\epsilon\epsilon^{ab}_{\phantom1\phantom2c}\pi^i_bE^c_iU^{-1}_\epsilon=\epsilon^{abc}(-\delta_{db}\pi^i_{c}E^d_i+\frac{1}{2}\delta_{db}\epsilon_{c}\epsilon^fE^d_i\pi^i_f-\frac{1}{2}\epsilon_f\epsilon_{b}\pi^i_{c}E^f_i)\neq {\varphi'}_\epsilon^a,\label{Urot}
\end{equation}

which does not coincide with the rotation constraint in the new sector. However, by multiplying the expression (\ref{Urot}) times $\epsilon_{cdf}\epsilon^f$ and retaining the leading orders in $\epsilon_a$, it can be shown \cite{CM07} that
the conditions $\epsilon^{abc}(-\delta_{db}\pi^i_{c}E^d_i+\frac{1}{2}\delta_{db}\epsilon_{c}\epsilon^fE^d_i\pi^i_f-\frac{1}{2}\epsilon_f\epsilon_{b}\pi^i_{c}E^f_i)\psi'=0$ and $\varphi_\epsilon^a\psi'=0$ are equivalent.

This result completes the proof that $U$ maps physical states between $\bar{\chi}=0$ and $\bar{\chi}=\epsilon$ sectors. Hence, even though the Lorentz frame is fixed before the quantization, nevertheless \emph{the boost symmetry can be realized on a quantum level}. 
 
Therefore, the gauge fixing of boosts does not imply any violation of the full Lorenz symmetry on a quantum level, in a second order formulation for the gravitation field.

Main prospectives of this work are\begin{itemize}{\item the investigation whether the introduction of matter fields modifies this picture,} {\item a reformulation in terms of first-order variables (as Ashtekar-Barbero-Immirzi ones), which can give new insights on the physical meaning of some ambiguities that arise after the quantization.}\end{itemize}


\end{document}